\documentclass[graybox, envcountchap]{svmult}

\usepackage{mathptmx}        
\usepackage{helvet}          
\usepackage{courier}         

\usepackage{makeidx}         
\usepackage{graphicx}        
\usepackage{multicol}        
\usepackage[bottom]{footmisc}

\makeindex             

\usepackage{amsmath}
\usepackage{amssymb}
\usepackage{multirow}
\newcommand{\apj}{ApJ}

\newcommand{\apjl}{ApJL}
\newcommand{\mnras}{MNRAS}
\newcommand{\aap}{A\&A}

\newcommand{\ssr}{SSRv}
\newcommand{\apss}{Astrophysics and Space Science}

\newcommand{\eqb}{\begin{equation}}  
\newcommand{\eqe}{\end{equation}}			

\newcommand{\rl}{r_{\rm L}}

\newcommand{\figurefour}{
\begin{figure}
\scalebox{0.9}{
\includegraphics{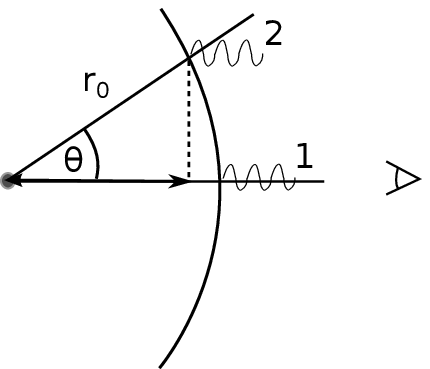} 
}
\caption{\small Pulsed emission from the wind. Photons 1 and 2 are emitted at the same time from the current sheet (modelled as a thin shell) when it passes the radius $r_0$. A distant observer will measure a time delay between the photons. If this time difference is smaller than the time between two subsequent current sheets crossing the sphere $r_0$, the emission from these sheets will appear as pulsed. }
\label{figfour}
\end{figure}
}

\newcommand{\figurefive}{
\begin{figure}
\scalebox{0.6}{
\includegraphics{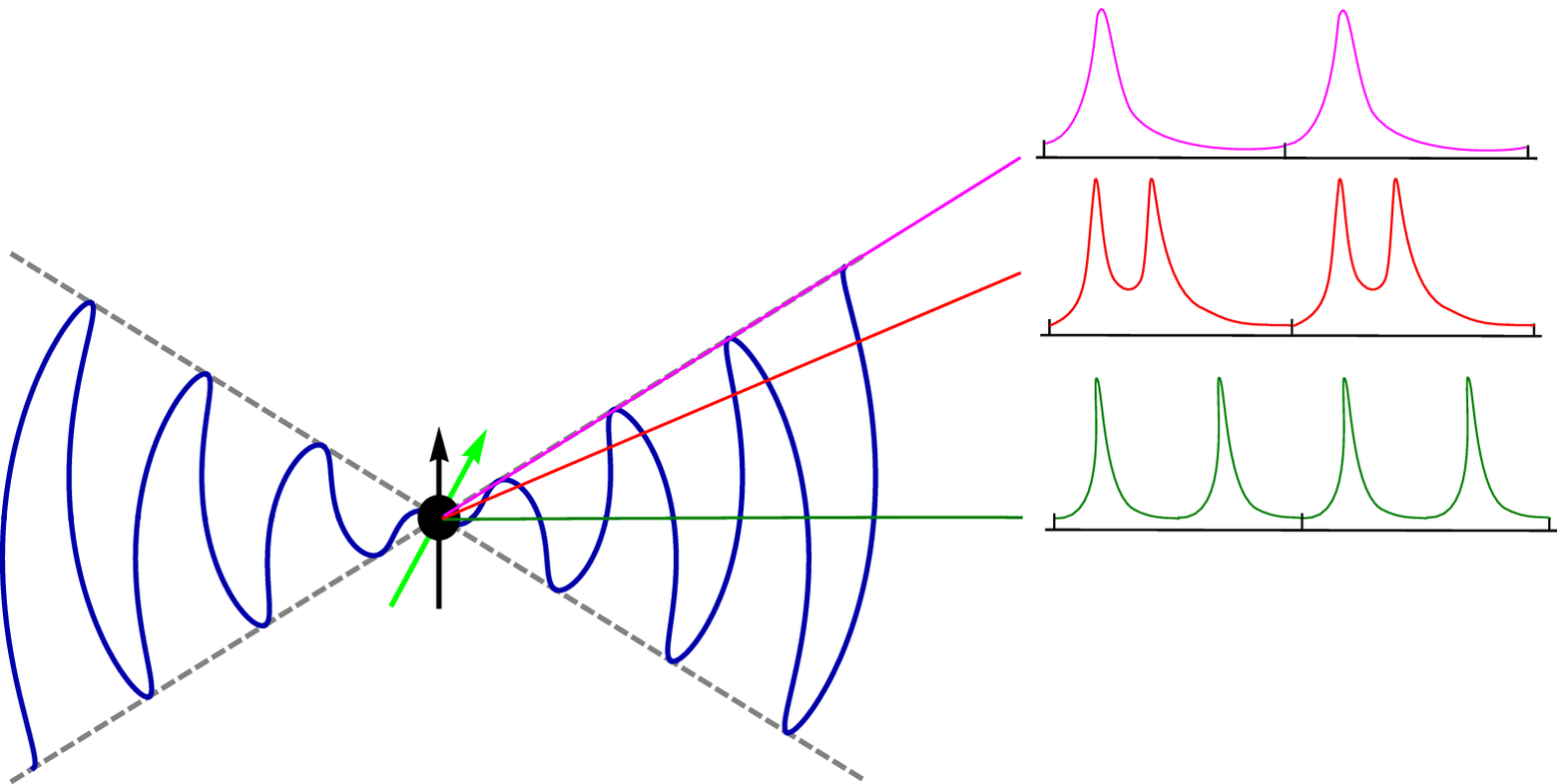} 
}
\caption{\small Lightcurves from the wind. A distant observer can detect a pulse of emission when  the  expanding current sheet passes the radius $r_0$ along his line of sight. The emissivity of this sheet quickly diminishes afterwards. Depending on the viewing angle, the observer can detect up to two pulses per rotational period of a pulsar.}
\label{figfive}
\end{figure}
}

\newcommand{\figureamano}{
\begin{figure}
\scalebox{0.5}{
\includegraphics{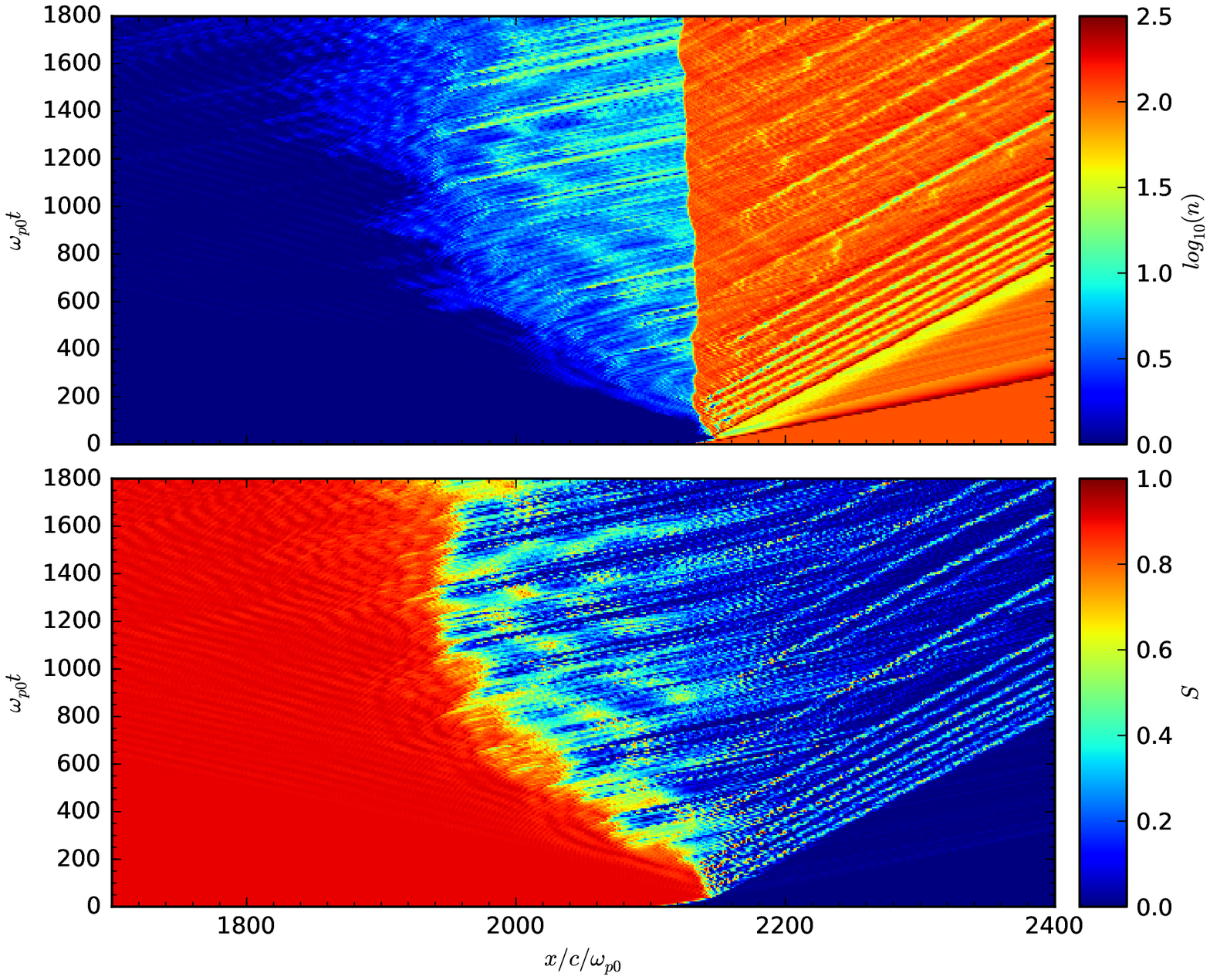} 
}
\caption{\small Two-fluid simulations carried out by Amano \& Kirk \cite{2013ApJ...770...18A} showing formation (space versus time) of an electromagnetic precursor, when a magnetic shear with the frequency exceeding the local plasma frequency is launched against a shock. The top and bottom panels show the proper density and Poynting flux, respectively. In the precursor region the Poynting flux becomes completely dissipated.}
\label{figamano}
\end{figure}
}

\newcommand{\figuresheet}{
\begin{figure}
\scalebox{0.4}{
\includegraphics{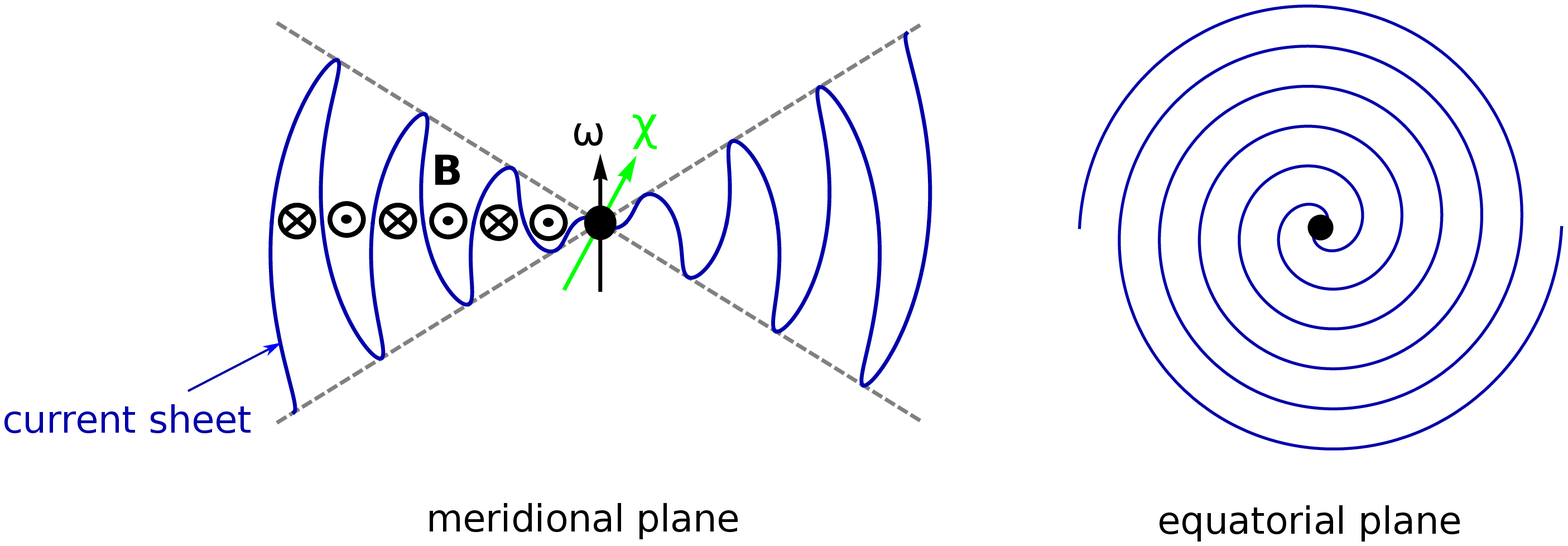} 
}
\caption{\small Striped wind for an oblique rotator with the angle $\chi$ between the magnetic and the rotational axes. In the meridional plane the current sheet (blue curve) is corrugated and separates the stripes of opposite magnetic polarity. The dominant component of the magnetic field is toroidal. In the equatorial plane the current sheet develops an Archimedean spiral.}
\label{fig_cs}
\end{figure}
}

\newcommand{\figuresironi}{
\begin{figure}
\scalebox{0.65}{
\includegraphics{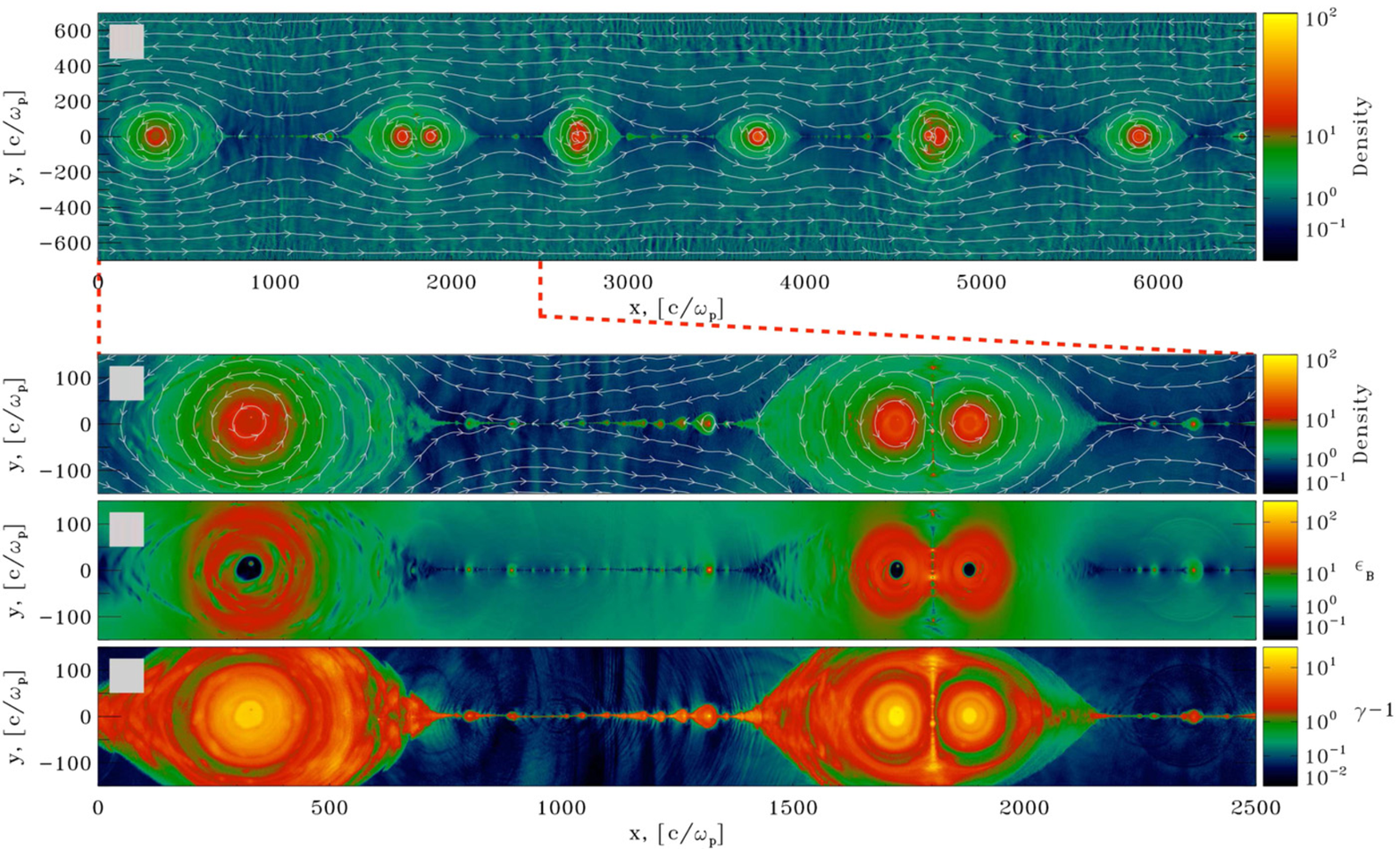} 
}
\caption{\small The structure of the reconnection layer, from PIC simulations of Sironi \& Spitkovsky \cite{2014ApJ...783L..21S}. First two panels show the particle density, the third panel shows the magnetic energy, and the bottom panel -- kinetic energy per particle.}
\label{figcs}
\end{figure}
}

\newcommand{\figuredistrib}{
\begin{figure}
\scalebox{0.9}{
\includegraphics{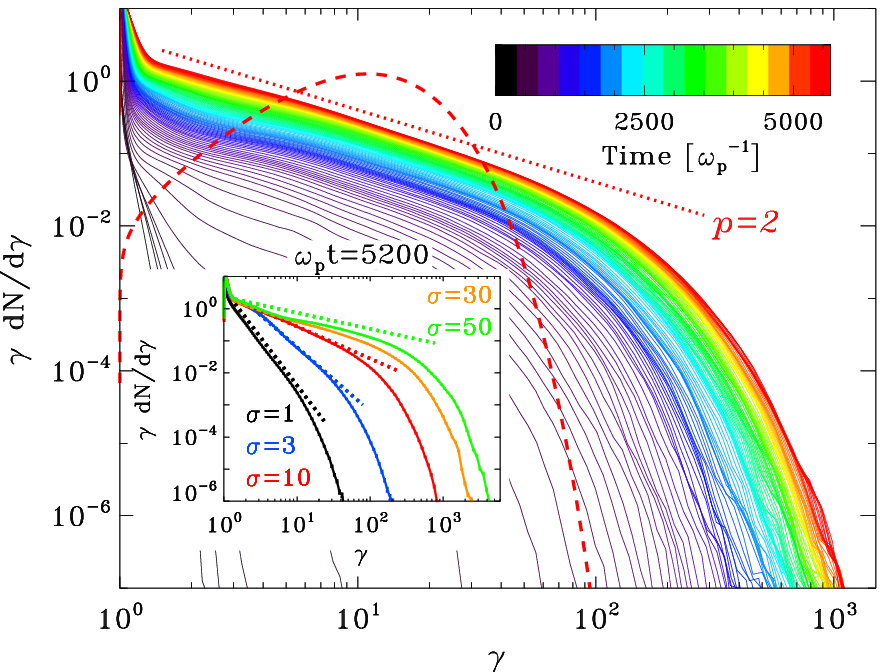} 
}
\caption{\small Evolution of the particle energy spectrum during reconnection in the plasma with magnetisation $\sigma=10$, from PIC simulations of Sironi \& Spitkovsky \cite{2014ApJ...783L..21S}. 
At late times the spectrum is a power law. 
The dependence of the spectrum on the magnetisation is shown in the inset; the dotted lines refer to power-law slopes of
−4, -3, -2, -1.5 from black to green.}
\label{figspectr}
\end{figure}
}

\begin{document}

\frontmatter

\tableofcontents

\include{cblist}

\mainmatter

%
%
%

%
%
%
%
%
%
%

\title{Pulsar striped winds}
\author{Iwona Mochol}
\institute{I.~Mochol \at Observatoire Astronomique de Strasbourg, 11 Rue de l'Universit\'e, F-67000 Strasbourg, France \\ Institute of Nuclear Physics, Polish Academy of Sciences, ul.~E.~Radzikowskiego 152, 31-342 Krak\'ow, Poland, \email{iwona.mochol@astro.unistra.fr}}
%
%
\maketitle

\abstract*{According to magnetohydrodynamic (MHD) models, the rotational energy of a rapidly spinning neutron star is carried away by a relativistic wind and deposited at a large distance, in the nebula, downstream of the wind termination shock. The energy transport in the outflow is mediated by Poynting flux, but it is not clear how the energy stored in the fields is transferred into the energized population of emitting particles. 
The most plausible dissipation mechanisms are thought to be related to the \lq\lq striped\rq\rq\  structure of the wind, in particular, to the existence of a current sheet, prone to reconnection events.  
In this model the current sheet is a natural place for 
internal dissipation and acceleration of particles responsible for pulsed, high-energy emission. Moreover, reconnection is a promising scenario for explaining annihilation of fields at the shock and conversion of their energy into the kinetic energy of  particles. The shock structure, however, is likely to differ in the low-density plasmas, in which non-MHD effects intervene. In this regime, the striped wind can dissipate its energy via an electromagnetic precursor of the shock.}

\abstract{ 
According to magnetohydrodynamic (MHD) models, the rotational energy of a rapidly spinning neutron star is carried away by a relativistic wind and deposited at a large distance, in the nebula, downstream of the wind termination shock. The energy transport in the outflow is mediated by Poynting flux, but it is not clear how the energy stored in the fields is transferred into the energized population of emitting particles. 
The most plausible dissipation mechanisms are thought to be related to the \lq\lq striped\rq\rq\  structure of the wind, in particular, to the existence of a current sheet, prone to reconnection events.  
In this model the current sheet is a natural place for 
internal dissipation and acceleration of particles responsible for pulsed, high-energy emission. Moreover, reconnection is a promising scenario for explaining annihilation of fields at the shock and conversion of their energy into the kinetic energy of  particles. The shock structure, however, is likely to differ in the low-density plasmas, in which non-MHD effects intervene. In this regime, the striped wind can dissipate its energy via an electromagnetic precursor of the shock.}

\section{Introduction}
\label{sec:1}

\subsection{General picture}

According to current models the entire post-supernova system is tightly coupled. 
The central engine -- a rapidly spinning neutron star -- is responsible for creation of particles, their acceleration and finally outflow in the form of a relativistic wind. The energy in the wind is transported by Poynting flux and released to the plasma at large distances from the star, in the nebula, where the energized particles radiate away the acquired energy and perform $p$d$V$ work on the surroundings. 
The global evolution of the system is usually modelled using relativistic MHD.
Although it allows us to get a glimpse on how the system is globally interconnected, we still do not understand the details of the coupling mechanisms and the processes that lead to the release of the magnetic energy.

The abundant pair plasma, transported by the wind, is thought to be created in electromagnetic cascades in the closest neighbourhood of the star, in the magnetosphere. The conventional boundary of this region is defined by the light cylinder of a cylindrical radius $\rl=cP/(2\pi)$ (where $P$ is the rotational period of a pulsar), at which the corotation velocity of the particles with the star would be equal to the speed of light. The corotating plasma is coupled to the magnetic field lines, which in the dipole-like geometry close within the light cylinder.  
The field lines which cross the light cylinder carry the currents that brake the central rotator via Lorentz forces. 

This rotational energy of the central neutron star is transported outwards by the wind, a mixture of plasma and electromagnetic fields. The outflow is terminated at the roughly standing shock, where its ram pressure becomes balanced by the pressure of the confining external medium. The shock is located at a large distance from the central star -- in the case of the Crab $r_{\rm shock}\sim 10^8 \rl$. At the shock the magnetic energy of the wind is randomised and the particles carried by the wind become efficiently accelerated via the Fermi I-order process. They are further injected into the nebula downstream of the shock, where they radiate in a synchrotron process, observed as diffuse emission from radio to X-rays; in gamma-rays a dominant process is the inverse-Compton (IC) scattering on the cosmic microwave background radiation and on the synchrotron photons produced in the nebula itself (synchrotron self-Compton (SSC) emission). 

Both the diffuse radiation from the nebula, as well as the high-energy pulses of emission observed from the central point-like source, provide the observational evidence that efficient mechanisms of particle acceleration are at work, challenging our understanding of the physics of relativistic flows. 
In the following we will concentrate on one particular underlying question: how and where the enormous magnetic energy carried by the wind is dissipated and transferred to the energised population of radiating particles?

\subsection{Problems}
 
The energy of the central rotator is transported over long distances, and, therefore, the wind zone proves to be an important ingredient in the global description of the system. On the other hand, very little direct information is provided by observations, in which the wind region appears underluminous 
\cite{2000ApJ...536L..81W}. This is related to the fact that the plasma in the wind is expected to be cold and, therefore, not producing any emission. However, if due to some internal dissipation the plasma were heated, the relativistic beaming and the radial dependence of the emissivity would cause the emission to appear as point-like \cite{2009ASSL..357..421K}.

The point-like emission from the relativistic, MHD pulsar wind would be seen by a distant observer as pulsed \cite{2002A&A...388L..29K}, because the internal dissipation in the wind, due to e.g. reconnection, is expected to occur abruptly in a thin, corrugated layer -- a current sheet (see Sect.~\ref{section:sw}). 
Reconnection regions are known to be powerful particle accelerators, and, therefore, this scenario provides a promising possibility to explain puzzling very high-energy $\gamma$-ray observations from the Crab \cite{2016A&A...585A.133A} and Vela \cite{velahess2, 2014ApJ...797L..13L} pulsars, as well as the millisecond pulsar J0614--3329 \cite{2016arXiv160408710X}. 
This is an alternative to the magnetospheric models, which associate the $\gamma$-ray pulses with cyclotron-self Compton scattering \cite{2013MNRAS.431.2580L}, synchro-curvature radiation \cite{2015MNRAS.449.3755V} and SSC emission inside the light cylinder \cite{2015ApJ...811...63H}.   

The pulsed emission, however, constitutes only a fraction of the total energy budget.  
On the global scale, the dissipation of the magnetic energy is believed to take place somewhere close to the termination shock. The properties of a shock, in particular its ability to dissipate the energy, strongly depend on the magnetisation of the upstream plasma. This magnetisation is quantified by the parameter $\sigma$. In a cold plasma it measures the ratio of the magnetic energy flux to the kinetic energy flux in the flow:
\eqb \sigma=\frac{B^2}{8\pi N mc^2 \Gamma}, \eqe
where $B$ is the strength of the magnetic field, $N$ is the plasma density in the laboratory frame and $\Gamma$ is the Lorentz factor of the flow. 
The magnetisation is thought to be large close to the light cylinder, where the wind is launched. 
However, Kennel \& Coroniti \cite{1984ApJ...283..694K} have shown that the observational limits on the magnetisation can be put downstream of the termination shock, in the nebula, where it must be small. It seems, therefore, that the transition between large- and small-$\sigma$ regimes requires some magnetic dissipation in the wind zone or at the shock itself. The mechanism responsible for this dissipation is unknown, with the magnetic reconnection being one of the most important candidates.

In the following we review the physics of relativistic pulsar winds and the dissipation processes that can lead to emission observed from those outflows. We start by considering their launching from the magnetospheres in Sect. \ref{section:launch}, and we discuss their \lq\lq striped\rq\rq\ structure in Sect. \ref{section:sw}. Further, in Sect. \ref{section:why}, we summarise the arguments in favour of the wind model of pulsed high-energy emission. 
In Sect. \ref{problem1} we consider a dissipation mechanism -- relativistic reconnection, which is likely to set in in the current sheet of the wind; a population of particles accelerated in the course of reconnection can be responsible for the pulses of emission, which would be, therefore, a direct probe of the physics in the current sheet. 
In Sect. \ref{section:implic} we discuss the implications of this model for the gamma-ray spectra of pulsars and its predictions for future observations. Finally, in Sect. \ref{section:shocks}, we describe the structure of shocks of striped winds and the $\sigma$-problem -- a failure to describe the conversion of the Poynting flux into the particle energy flux before the wind reaches the termination shock. A solution for this problem is still not clear, but we discuss two scenarios that have been proposed, namely the driven reconnection of wind stripes at the shock and the generation of an electromagnetic dissipative precursor to the shock.

\section{MHD picture: striped winds}

\subsection{Launching from a magnetosphere}
\label{section:launch}

Understanding of pulsar wind launching requires the knowledge
of (1) the magnetic field topology, (2) the location of particle acceleration regions, and (3) the mechanisms
of supplying charges to the outflow. The latter are closely linked to radiation processes,
which in the strong-field regime often change their character or become dominated by exotic phenomena, resulting from energy quantization in the direction perpendicular to the field (Landau states) or the
nonconservation of transverse photon momenta which can be absorbed by the field \cite{1991Sci...251.1033H}.

The magnetic field is usually modelled as a rotating dipole (strongly distorted close to the light cylinder), but higher multipoles may also contribute \cite{1982ApJ...254..713B, 2001ApJ...550..383G, 2015MNRAS.450..714P}. The strength of the magnetic field can be measured in two ways: (1) from the pulsar spin down, assuming that it results from the dipole emission, (2) directly from the
energies of cyclotron absorption lines in the X-ray spectra of neutron stars (for discussion see e.g. \cite{1991Sci...251.1033H}). 
These measurements imply that the magnetic fields at the stellar surface are very strong $B_*\gtrsim 10^{12}$~G.

A rotating neutron star, endowed with a dipole magnetic field, generates a quad\-ru\-po\-le electric field, in a
direct analogy to a unipolar inductor: like a conductor rotating in a magnetic field, charges
on the pulsar's surface are redistributed by the Lorentz force, which induces an electric potential
difference between the poles and the equator.
The generated electric field is thought to be strong enough to rip the charges off from the stellar surface and accelerate them along the magnetic field lines, because any transverse motion would be immediately suppressed by radiation of synchrotron photons. 
The accelerated charges emit curvature gamma-rays, which in the strong magnetic field can materialise into secondary $e^{\pm}$ pairs, further accelerated and radiating. 

The number of pairs created in these electromagnetic cascades is not clear. 
The critical Goldreich-Julian density $N_{\rm GJ}$ defines the number of charges that are necessary to screen out the component of the electric field parallel to the magnetic field:
\eqb N_{\rm GJ}=\frac{\mathbf{B}\cdot\boldsymbol{\rm \omega}}{2\pi e  c}\frac{1}{1-(r/\rl)^2\sin^2\theta}, \label{nGJ}\eqe
where $\mathbf{\rm \omega}$ is the angular velocity of the pulsar. 
The density of pairs $N$, created in the magnetosphere, is usually expressed in terms of the pair multiplicity 
\eqb \kappa=N/N_{\rm GJ}. \eqe 
If the pair injection in the magnetosphere is somewhat below $N_{\rm GJ}$, the pulsar magnetosphere is described as the \lq\lq electrosphere\rq\rq, in which a charge-separated plasma is confined to domes above the poles and a differentially rotating equatorial disc \cite{1985MNRAS.213P..43K,1985A&A...144...72K, 2002A&A...384..414P, 2004IAUS..218..357S, 2015MNRAS.448..606C}. 
If the beam of primary particles ripped off from the stellar surface initiates prolific cascades, a pair plasma in the magnetosphere is abundant $\kappa\gg 1$. 
Simulations of electromagnetic cascades suggest multiplicities in the range $\kappa \sim 10^3-10^5$ (see \cite{2015ApJ...810..144T} and references therein). 
A constraint on the multiplicity $\kappa \sim10^6$ is implied by the models of emission from the Crab Nebula, under the assumption that all the radiating particles are being constantly supplied by the pulsar and that these particles are in energy  equipartition with the magnetic field \cite{1974MNRAS.167....1R,1996ApJ...457..253D,2002ASPC..271...99G} (for discussion of the radio brightness see also \cite{2003MNRAS.345..153L,2009ASSL..357..421K}). Modelling of other nebulae in general suggests $\kappa\sim10^5$ \cite{2011MNRAS.410..381B}.
 
Such large values of multiplicity imply that the magnetospheric plasma easily screens the parallel component of the electric field. However, to maintain the activity, a pulsar has to sustain the constant or at least intermittent replenishment of pairs that outflow with the wind. Current theories postulate that the pair creation processes take place in limited regions of an unscreened component of the electric field -- so called vacuum gaps.   
Their location is model-dependent: they are associated with 
the polar caps (located near the stellar surface close to the magnetic axis of the pulsar) \cite{1971ApJ...164..529S, 1975ApJ...196...51R, 1978ApJ...225..226H, 1982ApJ...252..337D}, outer magnetosphere \cite{1986ApJ...300..500C, 1995ApJ...438..314R}, and the region along the last open field line \cite{1979ApJ...231..854A,2003ApJ...588..430M, 2004ApJ...606.1143M,2003ApJ...598.1201D}. 
Lyubarsky \cite{1996A&A...311..172L} suggested that the pairs can be produced in the current sheet just outside the light cylinder, where the magnetic field in the sheet is strongest. This scenario is observed in kinetic Particle-In-Cell (PIC) simulations of pulsar magnetospheres with low obliquity, whereas the pulsars with high obliquity tend to create pairs rather in polar cap regions \cite{2015ApJ...801L..19P}. 

The existence of gaps, although important for understanding of how pulsars really work, is usually neglected in the modelling of the global plasma dynamics. In particular, the MHD, fluid description of the plasma is thought to be valid.  
Since the energetics is dominated by electromagnetic fields, the particle 
inertia is negligible in the first approximation, but the particles are still assumed to carry charge and currents. 
This zero-inertia limit of MHD is called force-free electrodynamics (FFE). 
The equations for a force-free magnetosphere of an aligned rotator were obtained by Michel
\cite{1973ApJ...180L.133M}, as well as Scharlemann \& Wagoner \cite{1973ApJ...182..951S}. 
The plasma, rigidly corotating with the star, is assumed to have infinite conductivity, such that 
in the comoving frame the electric field vanishes; after transformation into the laboratory frame, this implies the frozen-in condition between the plasma and the magnetic field 
\eqb \mathbf{E}+(1/c)\mathbf{v}\times \mathbf{B}=0. \label{frozen-in} \eqe  
Corotation of a plasma with the star is possible only within the light cylinder. 
Since the particles cannot corotate with superluminal speed, the magnetic field lines that would cross the light cylinder are bent backwards and they open beyond $\rl$,  
even in the case of an aligned rotator (and in contrast with an aligned rotator in vacuum). 
As explained by Michel \cite{1973ApJ...180..207M}, in vacuum, without the plasma which ensures the frozen-in condition, the magnetic field of such an aligned rotator is static, whereas in the presence of a  plasma, the magnetic field is \lq\lq rotating\rq\rq. As a consequence, there is a nonvanishing Poynting vector $\mathbf{S}=\mathbf{E}\times \mathbf{B}$, which describes the transport of energy; the closer to the light cylinder, the more the corotation speed increases, and so does the magnitude of the Poynting vector. Ultimately, beyond $\rl$, it would exceed the energy density that can be transported with the speed of light. Thus, the field lines become open by the centrifugal force exerted by the corotating energy density of the electromagnetic fields. 

For an oblique rotator, with the angle $\chi$ between the magnetic and rotational axes, the presence of the plasma increases the torque when compared to a rotator in vacuum. The latter emits dipole radiation at the expense of the rotational energy, and the luminosity is given by:
\eqb L_{\rm vac}=L_{0}\sin^2\chi, \eqe
where $L_0=B_*^2\omega^4r_*^6/(6c^3)$, $B_*$ is the strength of the magnetic field at the stellar equator, $r_*$ is the stellar radius. 
The plasma-filled magnetosphere, on the other hand, spin downs according to
\eqb L_{\rm FFE}\approx \frac{3}{2}L_{0}(1 +\sin^2\chi) \eqe
\cite{2006ApJ...648L..51S,2012MNRAS.424..605P}. 
In comparison to the vacuum case, the topology of the magnetic field lines in the force-free description is similar inside the light cylinder, but exhibits the formation of a current sheet outside this distance \cite{2012MNRAS.424..605P}. 

The analytical studies of pulsar magnetospheres of oblique rotators are nowadays extended to the simulations: 3D force-free \cite{2006ApJ...648L..51S, 2012MNRAS.424..605P}, resistive force-free \cite{2012ApJ...749....2K, 2012ApJ...746...60L}, relativistic MHD \cite{2006MNRAS.367...19K, 2013MNRAS.435L...1T}, and also PIC simulations \cite{2015ApJ...801L..19P}.  
In general relativity force-free simulations have been performed recently by P\'etri \cite{2016MNRAS.455.3779P}. 

An analytical two-fluid approach to describe the pulsar magnetosphere has been proposed by Petrova \cite{2015MNRAS.446.2243P}. In this solution, electron and positron fluids exhibit a velocity shift with respect to each other, which can underlie the instabilities and production of waves, convertible into the radio emission from pulsars. Interestingly, in this model the plasma conductivity in the magnetosphere is finite due to inertial effects, which provides a physical justification for the resistivity, considered only phenomenologically as a free parameter in the resistive force-free simulations \cite{2012ApJ...749....2K,2012ApJ...746...60L}.    

\subsection{The striped wind structure}
\label{section:sw}

An exact force-free monopole solution for an outflow of an aligned rotator was found by Michel \cite{1973ApJ...180L.133M} and it was generalized for an oblique rotator by Bogovalov \cite{1999A&A...349.1017B}. 
In the pulsar models, the magnetic field is believed to change smoothly from the dipolar geometry inside the light cylinder to the monopolar, radial geometry at infinity \cite{1973ApJ...180..207M}. However, this change of magnetic field topology, captured numerically, has not been described analytically.  
Recently though, the exact dipole solution for an aligned rotator has been reported and discussed by Petrova \cite{2016arXiv160807998P}.

In the limit of negligible
particle inertia, a monopole configuration is radial in the meridional plane, 
and
in the equatorial plane these lines develop a perfect Archimedean spiral
$r_s = \rl \phi$ 
regardless of a field topology inside the light cylinder. Globally, more appropriate is a split-monopole configuration, where two half magnetic monopoles of opposite polarity are joined together 
in the equatorial plane. This change in magnetic field direction must be, according to Amp\`ere's law,
accompanied by formation of a current sheet, within which the magnetic field vanishes and the
pressure, necessary to keep the equilibrium, is supported by a hot plasma. 
In the oblique case, a current sheet oscillates around the equatorial plane
as the pulsar rotates, connecting the equator with field lines of opposite polarity every half
a period (see Fig.~\ref{fig_cs}). This corrugated current sheet far from the light cylinder can be approximated by spherical shells, separating the stripes of magnetized plasma with opposite magnetic polarity -- this is the structure of a so-called striped wind \cite{1990ApJ...349..538C, 1994ApJ...431..397M}. 
A solution for the wind with analytically prescribed current sheet and a radial velocity, has been recently reported \cite{2013MNRAS.434.2636P}.

\figuresheet 

In the split-monopole solution the Poynting flux has a maximum value at the equator \cite{1973ApJ...180L.133M}. Most of the energy, therefore, is carried within the wedge of the striped wind, defined by the spherical coordinate $\theta$, $\pi/2-\chi<\theta<\pi/2+\chi$, where $\chi$ is the angle between the magnetic and rotational axes of the pulsar (obliquity).
In the equatorial plane the stripes have the equal width, hence the phase-averaged magnetic field vanishes. This is not the case for the higher latitudes, for which 
a phase-averaged magnetic field increases with the latitude. It reaches the maximum value at the surface defined in spherical coordinates by $\theta=\pi/2\pm\chi$. This surface passes through the cusps of the current sheet. Above it, in the polar region of the wind, the total magnetic field has only one polarity.

The radial dependence of the magnetic field components can be obtained from the magnetic flux conservation. 
Integrating over the volume of a sphere of a radius $r$, one gets from the Gauss
theorem
\eqb 
B_r\propto r^{-2}. \label{eqbr}
\eqe
Similarly, as the number of field lines in the sectional area is constant,
\eqb
B_{\phi}\propto r^{-1} \label{eqbphi}
\eqe 
(for a detailed discussion see e.g. \cite{1994ApJ...431..397M, 2009ASSL..357..421K}). 
Close to the light cylinder the radial and poloidal component of the magnetic field are comparable, 
but the poloidal component, which in this solution is purely radial, decreases much faster than the
toroidal one. Observations of synchrotron radiation close to the termination shock, in particular its uniform
linear polarization, are consistent with the toroidal structure of the magnetic field, expected in the wind at large distances from the pulsar.

The radial wind is believed to emerge from the magnetosphere as submagnetosonic. 
In the force-free solution the fast magnetosonic (FMS) point is located at infinity, and the outflow accelerates with the distance $\Gamma\propto r$ \cite{1977MNRAS.180..125B} at the expense of the magnetic energy. As a result, the wind described by the FFE arrives at the termination shock as kinetically dominated \cite{2002ApJ...566..336C}. However, the FFE model neglects the particle inertia, which, by definition at the FMS point become important (propagation of the FMS wave is an interplay between particle inertia and magnetic
tension, implying that the former cannot be neglected). The FMS speed in a highly magnetized plasma is relativistic, defined by the Lorentz factor $\Gamma_{\rm fms}\approx\sqrt{\sigma} \gg 1$. For large-$\sigma$ flow, the FMS point should be crossed somewhere close to the light cylinder. Further out the wind becomes supermagnetosonic, described by MHD rather than FFE equations. An MHD relativistic wind is moving ballistically, with almost a constant, relativistic speed, and it stays strongly magnetically dominated up to the large distances (for a discussion of this topic see \cite{2009ASSL..357..421K}).

\section{Reconnection and pulsed emission from the wind}

\subsection{Why the wind model?}
\label{section:why}

The striped wind is a promising site of emission of high-energy (optical to gamma-ray)
nonthermal photons observed from pulsars, an idea proposed by \cite{2002A&A...388L..29K}.  
The difference between the wind and the magnetospheric scenarios lies in assumptions about the physics of production of the pulsed emission.
The wind model offers an analytic description of the magnetic field structure, and the peaks in a lightcurve appear as a result of the strong beaming (since the wind is relativistic) of the radiation produced in a thin layer -- a current sheet. This layer is naturally prone to dissipative processes, in particular to reconnection.  
On the other hand, in the magnetospheric models the peaks in a light curve
are caused by the overlap of emission from different
regions in the magnetosphere (caustics) (see e.g. \cite{2010ApJ...715.1270B}); this overlap is sensitive
to both the definition of gaps and the exact geometry of the magnetic field -- firstly, because the
geometry of the field defines how the effects of aberration of light and light
travel delay add together to form the caustics; 
secondly, because the geometry of the field close to the light cylinder determines
the shape of the polar cap on the star, which in turn controls the shape of the radiation region \cite{2010ApJ...715.1270B}. This requires a global modelling of the pulsar magnetospheres, and presumably can be done only via numerical simulations.

Moreover, the reconnection in the wind is expected to be a very powerful particle accelerator, whereas details of acceleration processes (and accelerating fields) that lead to magnetospheric emission are strongly model-dependent. 
Extreme conditions of particle acceleration are indeed required to explain 
recent observations of pulsars above a hundred of GeV \cite{2016A&A...585A.133A,velahess2, 2016arXiv160408710X}.

To understand how the pulses from the wind appear in the lightcurves, we assume that the corrugated current sheet can be described as subsequent shells, which radiate after crossing a sphere of a radius $r_0$ (see Fig.~\ref{figfive}). A distant observer will notice a peak of emission from the shell that has crossed the radius $r_0$ along his line of sight, followed by a quick decline of the emissivity as the sheet propagates outwards, due to adiabatic cooling and radial dependence of magnetic field (Eq. \ref{eqbr} and \ref{eqbphi}). 
Depending on the viewing angle, the observer can detect up to two pulses per rotational period of a pulsar -- 
(1) two pulses with changing separation if his line of sight lies in the wedge $\pi/2-\chi<\theta_{\rm obs}<\pi/2+\chi$; (2) one pulse, if the line of sight is close to $\theta_{\rm obs}\approx\pi/2\pm\chi$; (3) no pulse if $\theta_{\rm obs}<\pi/2-\chi$ or $\theta_{\rm obs}>\pi/2+\chi$.  
\figurefive

\figurefour
The pulses of radiation, if emitted too far from the pulsar, would be, however, smeared out. Let the observer be located is at the distance $D$ from the pulsar. In Fig.~\ref{figfour} photons 1 and 2 are emitted at the same time from the shell passing the radius $r_0$ (for simplicity we assume that the shell is spherical, which is true in the first approximation far from the light cylinder). The first photon arrives at the observer after time: $t_1=(D-r_0)/c$, whereas the second photon arrives after time $t_2=(D-r_0\cos\theta)/c$. Time time difference between arrival of the photons is therefore 
$\Delta t=r_0\left(1-\cos\theta\right)/c\approx r_0/(2\Gamma^2c)$,  
where for ultrarelativistc flow with the Lorentz factor $\Gamma\gg1$, $\theta\approx 1/\Gamma$. 
If this time difference is smaller than the time between two subsequent current sheets crossing the sphere $r_0$, the emission from these sheets will appear at the observer as pulsed. At the equator the shells are separated by half of the wavelength of the wind $\lambda/2=\pi \rl$. Thus, the condition for the maximal distance at which the emission emitted still will appear as pulsed can be estimated as:
\eqb r_0 < 2\pi \rl \Gamma^2 \eqe
For a detailed discussion and the calculations taking into account an exact form of the Archimidean spiral, see \cite{2016arXiv160804895P}.

It is worth mentioning that the pulse and off-pulse polarization features, modelled in the wind scenario for optical emission of the Crub pulsar, fit very well with this picture \cite{2005ApJ...627L..37P}. For more details see also \cite{2009ASSL..357..421K}.

If the pulsed emission originates in the wind, one expects that the spectrum of the received radiation should directly probe the physics of particle acceleration in the current sheet.

\subsection{Relativistic reconnection and particle acceleration in the wind}
\label{problem1}

In highly conducting plasmas, the magnetic energy can be released by the magnetic field reconfiguration near the singular regions, where the magnetic field changes direction and the localized currents introduce some non-ideal effects, breaking the frozen-in condition (\ref{frozen-in}). In the collisionless plasmas, these non-ideal effects include the electron kinetic/inertial terms in the Ohm's law, and microinstabilities, which on the macroscopic level play a role similar to the collisional resistivity. Rearrangement of the field topology occurs locally at so-called X-points. 

In the pulsar wind, magnetic reconnection is expected to set in in the current sheet, which separates the regions of oppositely polarized magnetic field. In highly magnetized plasma in the wind it proceeds in the relativistic regime, because initially the magnetic energy per particle is significantly greater than the particle energy. Thus, when the particles enter the reconnecting layer, they become accelerated to relativistic energies.  
While the steady-state description of relativistic reconnection has been studied analytically
\cite{2003ApJ...589..893L,2005MNRAS.358..113L}, numerical simulations (see \cite{2015SSRv..191..545K} and references therein) demonstrate that this process is essentially time-dependent; current layers are violently unstable to
the plasmoid (tearing) instability which leads to the continuous formation and ejection of multiple
secondary islands -- plasmoids (see Fig.~\ref{figcs}). 
\figuresironi

The whole process of particle acceleration depends on how the particles penetrate and interact with plasmons, which in turn merge and grow. In the full time-dependent picture the process of particle acceleration can only be captured by fully kinetic PIC simulations \cite{2001ApJ...562L..63Z,2004PhPl...11.1151J, 2007PhPl...14e6503B, 2014ApJ...783L..21S, 2014PhRvL.113o5005G, 2014ApJ...782..104C}.
The results suggest that the particles gain most of the energy when they are accelerated linearly by the electric field at X-points. 
When the tension force of the reconnected magnetic field advects the particles away from the X-point, they become trapped in magnetic islands, with a roughly isotropic energy distribution. 
The larger the size of a plasmoid, the more energetic particles it can confine. This provides a Hillas criterion for the Larmor radius of particles accelerated in the course of relativistic reconnection \cite{2016MNRAS.462...48S}. It has been reported \cite{2016MNRAS.462...48S} that the observed highest energy particles have Larmor radius of about 3\% of the system length (along the reconnection layer).

Integrated over the whole reconnection region, the energy spectrum of particles is generally observed to be a power law, with a very hard index $s<2$ for magnetisation $\sigma>10$ \cite{2014ApJ...783L..21S}, and being the harder the larger is the magnetisation parameter (see Fig.~\ref{figspectr}). 

\figuredistrib

The most energetic particles are the ones that interact closer to the centre of the X-point, because they are less prone to be advected away by the reconnected field and, therefore, they stay longer in the acceleration region \cite{2014ApJ...783L..21S}. 
The maximum energy to which a power-law particle distribution extends, as well as the shape of the high-energy cutoff in this distribution, both depend on the interplay between the acceleration process and the losses that the particles experience. 
The particle spectra obtained from PIC simulations by Werner et al. \cite{2016ApJ...816L...8W} exhibit either exponential or super-exponential behaviour. The exponential cutoff is more appropriate if the maximum energy is limited by the total available magnetic energy per particle, whereas the super-exponential shape better fits the results if the acceleration is limited by the particle escape from the acceleration region. 

In pulsars there are three important timescales that play a role in shaping a particle distribution during the acceleration and radiation processes. The first one is the timescale of acceleration of a particle to a Lorentz factor $\gamma'$ in the reconnection electric fields $E'=\tau B'$, where $\tau$ is the reconnection rate, and the prime denotes the quantities considered in the wind comoving frame:
\eqb t'_{\rm acc}=\gamma' mc/(eE'). \eqe
The second timescale is defined by radiative cooling, given by usual synchrotron losses of a particle with a Lorentz factor $\gamma'$: 
\eqb t'_{\rm rad}=6\pi mc^2/(\sigma_{\rm T}c\gamma' B'^2). \eqe 
The third timescale can be related to the particle escape from the acceleration region, when its gyroradius $r_{\rm g}=\gamma'mc/(eB')$ becomes comparable to the size of the confining region. In this case the highest energy particles are defined by the Hillas criterion \cite{2016MNRAS.462...48S}.

A particle is able to reach the radiation reaction limit when the timescale of the acceleration is equal to the timescale of radiative cooling in the wind comoving frame; in this case:
\eqb \gamma'_{\rm rad}=\left(\frac{6\pi e\tau}{\sigma_{\rm T}B'}\right)^{1/2}. \label{gamma_rad} \eqe
In fact, at the X-points the particles can reach even higher energies \cite{2004PhRvL..92r1101K,2012ApJ...746..148C,2014ApJ...782..104C}, because the radiation reaction force vanishes in the direct field acceleration. As a consequence, the particle distribution is not cut-off sharply at (\ref{gamma_rad}), but this energy defines where the cut-off starts in the distribution. 

The escape from the acceleration region will occur when the particle gyroradius becomes a fraction $\zeta$ of a current sheet thickness $\Delta_{\rm cs}=2\pi \beta c/\omega$ (the confining region could be similar to the size of the largest plasmoids, according to the results of simulations, it is a few percent of the system length -- here $\sim \rl$ \cite{2016MNRAS.462...48S}), which implies the limiting energy:
\eqb \gamma'_{\rm esc}=\frac{\Gamma \zeta (\Delta_{\rm sc} 2\pi c) eB'}{mc^2\omega}. \label{gamma_esc} \eqe

It can be expected, therefore, that in large current sheets with large plasmoids the particles will stay confined and accelerated until they reach the radiation reaction limit. On the other hand, in smaller systems, they will be mainly escaping, before reaching the radiation reaction limit.

It is interesting to consider two cases in which the limiting energy is defined by:
\begin{itemize}
\item the radiative cooling, and the particle distribution takes a form of a power-law with an exponential cutoff:
\eqb n(\gamma') \propto \gamma'^{-s}e^{-\gamma'/\gamma'_{\rm rad}}, \label{distr_rad} \eqe
\item the escape from accelerating/confining region. In this case the particle distribution is cut-off super-exponentially:
\eqb n(\gamma') \propto \gamma'^{-s}e^{-\gamma'^2/\gamma'^2_{\rm esc}}. \label{distr_esc} \eqe
\end{itemize}
This approach is, in fact, phenomenological, because in reality one should solve the continuity equation with a given acceleration process and radiative losses, in order to obtain the exact form of the particle distribution function. This, however, is a difficult problem, given, in particular, the complexity of the particle acceleration at reconnection sites.

\begin{figure}
\scalebox{0.9}{
\input{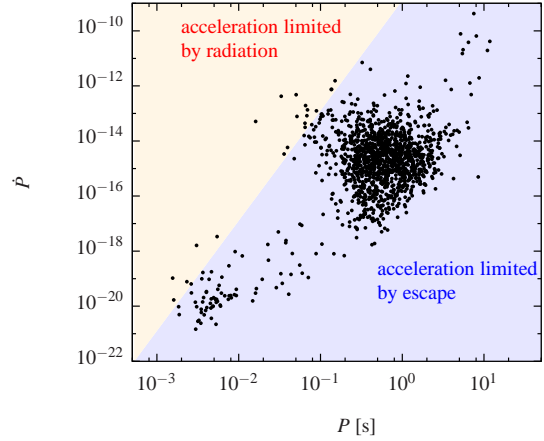} 
}
\caption{\small $P\dot{P}$-diagram showing the line (\ref{line}) which divides pulsars into two populations with different regimes of particle acceleration during reconnection in the wind.}
\label{figpp}
\end{figure}

The reconnection is assumed to set in in the wind at some distance $r=\hat{r}r_{\rm L}$ from a pulsar. Only a small fraction $\varepsilon_{\rm d}$ of the magnetic energy goes into acceleration of the plasma particles so that the process has some efficiency, but it does not alter the dynamics of the wind itself, and the rest of the energy stays either in the magnetic field -- for instance plasmons -- or in a thermal population of particles. 
With these assumptions the regime of the particle acceleration (radiative cooling/escape) for a given pulsar depends only on its parameters -- the spin down power and the period, given that $\gamma'_{\rm rad}\sim\gamma'_{\rm esc}$ when 
\eqb \left(\frac{\dot{E}}{10^{38} {\rm erg/s}}\right)^{3/2}\left(\frac{P}{0.01 {\rm s}}\right)^{-1}\sim 0.005 \left(\frac{r}{10 r_{\rm L}}\right) \left(\frac{0.1}{\zeta}\right)\left(\frac{0.1}{\Delta_{\rm cs}}\right) \left(\frac{\Gamma}{100}\right)^{3/2} \left(\frac{\varepsilon_{\rm d}}{0.01}\right)^{1/2}. \label{line} \eqe
This suggests that only the most energetic and the fastest pulsars exhibit acceleration occurring in the radiative cooling regime \cite{2015MNRAS.449L..51M}. When plotted on the $P\dot{P}$-diagram (see Fig.~\ref{figpp}) the dividing line separates two pulsar populations.

\subsection{Implications for gamma-ray spectra of pulsars}
\label{section:implic}

Radiative signatures of each acceleration regime are determined by the power-law index of the particle distribution, as well as the shape of the cutoff at the highest energies. 
The most important emission process in the current sheet is the synchrotron radiation, produced by the particles moving in the leftover and/or adjacent magnetic field. 
In addition, the SSC spectral component may become significant in the very high energy band, because the same population of leptons that emits synchrotron photons in the sheet, can also upscatter them to higher energies in the IC mechanism. This scattering proceeds in the Thomson regime when the electron Lorentz factor $\gamma$ and the photon energy normalized to the electron rest mass $\epsilon$ satisfy the condition $\gamma \epsilon\lesssim 1$. Thus, the soft photon field in the current sheet has to be sufficiently dense in order to make the scattering by the most energetic particles observable. 

The behaviour of the synchrotron spectrum at the highest energies can be calculated analytically from the synchrotron flux integral using the steepest descent method \cite{2015MNRAS.449L..51M}. The summary of the two considered cases is shown below in Table~\ref{tableregimes}.

\begin{table}
\centering
\normalsize
\begin{tabular}{c c c c c }
\hline \vspace{-5pt} \\ 
acceleration limit & \hspace{10pt} & radiation & \hspace{10pt} & escape \vspace{7pt}  \\ 
particle distribution & \hspace{10pt} & $n(\gamma)=n_0\gamma^{-s}e^{-\gamma/\gamma_{\rm rad}}$ & \hspace{10pt} & $n(\gamma)=n_0\gamma^{-s}e^{-\gamma^2/\gamma^2_{\rm esc}}$ \vspace{7pt} \\ 
asymptote of & \hspace{10pt} &
\multirow{2}{*}{$x^{1.3-(s+0.6)/3}e^{-1.9\,x^{1/3}}$}  & \hspace{10pt}  &
\multirow{2}{*}{$x^{1.3-(s+0.6)/4}e^{-2\,x^{1/2}}$}  \\
synchrotron spectrum & \hspace{10pt}  & &  \hspace{10pt} &\vspace{7pt} \\
SSC component & & detectable & & weak (not shown) \vspace{7pt} \\ \hline  
\end{tabular}
\caption{Summary of two regimes of particle acceleration in the wind.}
\label{tableregimes}
\end{table}

In order to model the emission from the wind the following procedure is adopted: (1) calculation of the synchrotron and SSC spectrum of an accelerated particle population (\ref{distr_rad}) and (\ref{distr_esc}) in the wind comoving frame, (2) transformation of the emission to the observer frame, (3) comparing the obtained spectrum of emission with the observed pulsed emission from a given pulsar, and finally (4) fitting the model to the data points in order to constrain $\Gamma$, $\hat{r}$, $s$ and $\varepsilon_{\rm d}$. 

\begin{figure}
\scalebox{0.8}{
\input{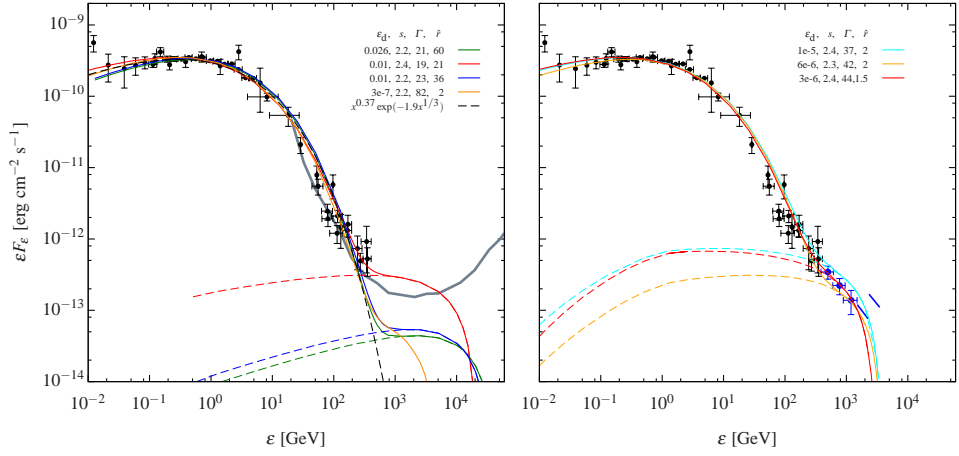}}
\caption{\small Left: Several fits to the Crab spectrum up to 400 GeV. Black points are the
data from \cite{2001A&A...378..918K,2010ApJ...708.1254A,2014A&A...565L..12A}. Thick gray line shows the predicted CTA sensitivity \cite{2013APh....43..287D}, black dashed line is the asymptote, 
plotted with $s=2.2$.
Dashed lines show SSC components, while solid lines are the total (synchrotron + SSC) spectrum. Right: Preliminary constraints on the model from the new observations from MAGIC \cite{2016A&A...585A.133A}, shown here by the blue points.}
\label{fig1}
\end{figure}

\begin{figure}
\scalebox{0.8}{
\input{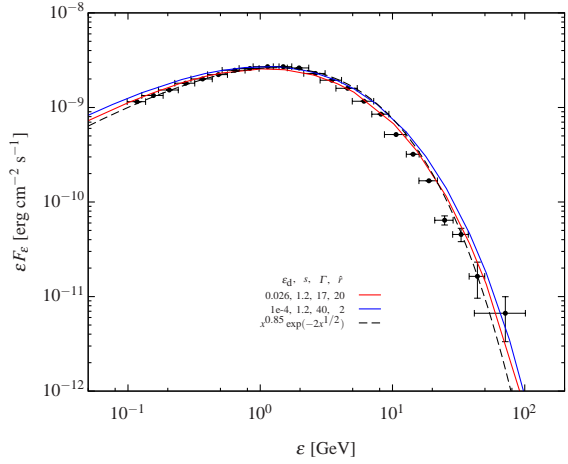}}
\caption{\small Several best fits to the synchrotron spectrum of the Vela. For a hard particle index, the SSC component is weaker by several orders of magnitude and it is not shown in the plot. Black dashed line is the asymptote, plotted with $s=1.2$. 
Black points are the data from Fermi-LAT \cite{2014ApJ...797L..13L}.}
\label{fig2}
\end{figure}

Two scenarios of acceleration, limited either by radiation reaction or the particle escape, are considered for an example of the Crab and Vela pulsars, respectively. The results are shown in Fig.~\ref{fig1} and Fig.~\ref{fig2}.

The emphasis should be put on the fact that fitting the spectrum to the data is not unique. 
If the dissipation efficiency $\varepsilon_{\rm d}$ is small, the density of radiating particles is low, and, therefore, to match the data points, their emission must be generated close to the light cylinder, where the magnetic field is stronger, and, in addition, the radiation has to be boosted by a large Lorentz factor of the wind. 
If $\epsilon_{\rm d}$ is larger, there are more accelerated particles, and, therefore, in order to not overproduce the synchrotron flux, the emission should be produced further from the light cylinder, where the magnetic field is weaker.

Interestingly, the difference between power-law indices $s$ in the two acceleration scenarios is close to 1, as expected for the radiatively cooling population with $s\sim2$ in comparison to the injected population with a very hard index $s\sim1$. That would be consistent with the model, in which  the cooling population is observed above the cooling break, and the escaping population is a real injection distribution.

Moreover, each of two considered regimes predicts different SSC component at TeV energies (compare Fig. \ref{fig1} with Fig. \ref{fig2}).
The maximum energy to which the synchrotron photons can be upscattered is given by
\eqb \epsilon_{\rm max}\approx 2\Gamma \gamma'_{\rm rad}mc^2\approx 3.6\,\, {\rm TeV} \left(\frac{\Gamma}{10}\right)^{5/2}\left(\frac{P}{0.01 {\rm s}}\right)^{1/2}\left(\frac{\varepsilon_{\rm d}}{0.01}\right)^{1/2}\left(\frac{\dot{E}}{10^{38} {\rm erg/s}}\right)^{-1/4} \eqe

The particle index changes the brightness of the SSC component, because it controls how many soft photons, available for upscattering (in the Thomson regime), are produced. For the steep particle index $s>2$ (like for the Crab), there are more low energy particles that can produce low-energy synchrotron photons; in this case SSC component is brighter. From the currently available data we can constrain the upper limit on this index to be $\sim2.4$, because steeper indices would overproduce the SSC emission. On the other hand, a fit to the Vela spectrum requires a much harder index $s<2$. In this case the SSC component is
weaker by several orders of magnitude (not shown in the
plot), because the energy density resides mainly in the highest energy particles, which do not produce enough low-energy synchrotron photons that can be upscattered to very
high energies.

For the Crab, the key to constrain the model and the pulsar wind parameters is partly given by the recent MAGIC results \cite{2016A&A...585A.133A}, which seems to prefer emission produced at smaller distances from the light cylinder. More accurate limits will be given by observations from the experiments HESS II and CTA.

\section{MHD and beyond: $\sigma$-problem and the structure of termination shocks of striped winds}
\label{section:shocks}

According to MHD models, the energy transported by the wind is released in the nebula downstream of the wind termination shock, where a broad-band  diffuse emission is produced. 
The Rankine-Hugoniot jump conditions at the shock have been solved by Kennel \& Coroniti \cite{1984ApJ...283..694K,1984ApJ...283..710K}. With the assumption that the particles are efficiently accelerated at the shock, they calculated the synchrotron emissivity downstream of the shock as a function of the upstream wind parameters. They found that MHD shocks in highly magnetized flows are very weak, implying that the magnetisation and the flow velocity practically do not change across them. If the flow is initially highly magnetized, it stays so after passing a shock without any significant energy dissipation.
Thus, in order to satisfy the observational constraints on the synchrotron emission, as well as the nonrelativistic expansion of the outer edge of the nebula, the plasma downstream, as well as upstream of the shock, has to be weakly magnetised.   
However,
according to theoretical pulsar models, the wind is highly
magnetised at its launching point and it stays so up to the termination shock. The reason is that a radial, large-$\sigma$ MHD flow
does not collimate as it propagates, and therefore it does not
convert the energy from the electromagnetic (EM) fields to the kinetic form, arriving at the shock still Poynting dominated.
Thus, it is not clear how and where the wind dissipates its EM energy to
the plasma, a puzzle known as the \lq\lq$\sigma$-problem\rq\rq. 
It has been shown that the phase-averaged, non-oscillating component of the fields in the
wind can dissipate in the bulk of the nebula
\cite{2013MNRAS.431L..48P}. Therefore the
$\sigma$-problem concerns mainly the dissipation of the wave-like
oscillating component of the fields. 

Lyubarsky \cite{2003MNRAS.345..153L} has
proposed a solution, in which the striped wind, due to interaction with the shock, becomes compressed and the stripes
dissipate the energy by the driven magnetic reconnection. This scenario provides not only the solution to the $\sigma$-paradox, but also explains the very hard spectrum of the nebula in the radio band as the synchrotron emission of the particles accelerated at the shock in the course of reconnection. 
PIC simulations \cite{2007A&A...473..683P,2011ApJ...741...39S} show that this mechanism operates also in 2D and 3D. 
The only caveat is that in order to reproduce the observed particle spectra,
a high plasma density has to be assumed $\kappa\sim10^8$, much higher than usually invoked in pulsar studies. 
In fact, in
plasmas of the assumed densities, reconnection in the wind
would start much earlier, before it arrives at the shock \cite{2003ApJ...591..366K}.

The question arises if there is a mechanism dissipating the magnetic energy in the regime of low plasma density.  
As pointed out by Usov \cite{1975Ap&SS..32..375U} and Michel \cite{1994ApJ...431..397M}, according to the mass continuity
$r^2 N v_r ={\rm const}$, the density of current carriers $N$, as measured in the laboratory frame, drops as $N\propto r^{-2}$,
 faster than the magnetic field $B_{\phi}\propto r^{-1}$ that they are required to maintain. 
As discussed by Coroniti \cite{1990ApJ...349..538C}, Michel \cite{1994ApJ...431..397M}, and also Lyubarsky \&
Kirk \cite{2001ApJ...547..437L}, 
the particles whose density decreases are forced to stream with higher
and higher drift-speed in order to satisfy the Amp\`ere's law in the flow, but, since the drift speed cannot reach $c$, the anomalous resistivity arises, which in
turn will trigger the magnetic reconnection. As a result, magnetic energy would be released into
heating of a plasma, which, however, will perform work
on the flow, leading to its acceleration \cite{2001ApJ...547..437L}.  
In a generic case, therefore, no significant dissipation will occur before the MHD wind arrives at the
shock, which must be then itself responsible for the wave dissipation.
On the other hand, if beyond a certain distance from the pulsar the flow becomes charge-starved, and therefore non-stationary, the displacement current in Amp\`ere's law cannot be neglected anymore. Usov \cite{1975Ap&SS..32..375U} and Melrose \& Melatos \cite{1996MNRAS.279.1168M} suggested that the MHD framework is not able to describe the whole physics of a diluted flow, which resembles more an EM wave in a plasma than an MHD wave.  
In this case one must refer to a more general, two fluid approach. Self consistent, analytic solutions of Maxwell and two-fluid equations have been discussed in numerous studies \cite{1971PhRvL..27.1342M,1974JPlPh..12..297C,1976JPlPh..15..335K,1984A&A...139..417A,2012ApJ...745..108A,2013ApJ...771...53M}. 

The propagation of EM waves in a plasma is possible only when their frequency $\omega_{\rm w}$ exceeds the proper plasma frequency $\omega_{\rm pl}$. If the amplitude of an EM is large $a=eB/(mc\omega)\gg1$, these waves can drive particles to relativistic Lorentz factors $\gamma\approx a\gg1$, and the condition of propagation is less restrictive. In the vicinity of the pulsar the propagation condition of a wave with a frequency of the rotating pulsar,
\eqb \omega_{\rm w}=\omega_{\rm pulsar} > \omega_{\rm pl}=\left(\frac{8\pi e^2 N}{\gamma m}\right)^{1/2}, \eqe
where $N=\kappa N_{\rm GJ}\rl^2/r^2$ and $N_{\rm GJ}$ is given by Eq.~\ref{nGJ}, translates to
\eqb r > r_{\rm crit}, \qquad r_{\rm crit} = \kappa \rl \label{propcond} \eqe

For most isolated pulsars the shock is located at a sufficiently large distance to allow the existence of the underdense region. 
For the Crab pulsar $\kappa\sim10^6$, and thus the shock located at $r_{\rm shock}\sim10^9 \rl$, satisfies the condition (\ref{propcond}). 
Since the EM waves become new eigenmodes in the system, they can be generated when the rarified wind interacts with the shock, forming a dissipative precursor. This precursor accelerates the particles to relativistic energies and 
transfers the energy from the fields into the plasma. 

\figureamano
The two-fluid simulations \cite{2013ApJ...770...18A} of this process have indeed demonstrated the formation of such precursors in diluted plasmas (see Fig.~\ref{figamano}). The authors simulated a circularly polarized MHD wave (magnetic shear), mimicking an MHD wind, launched against a shock in a low density plasma, i.e. whose plasma frequency is smaller than the frequency of the wave. When the wave hits the shock, it is  observed that EM waves are generated in front of the shock, and the system relaxes to an equilibrium in which an electromagnetic precursor is formed, and the Poynting flux of the initial wave is dissipated by the EM modes. As a consequence, a shock becomes essentially unmagnetized and thus potentially a very efficient Fermi accelerator. 
Recent studies \cite{giacchekirk} of the particle dynamics in the precursor indeed demonstrate its ability to energise particles via wave-particle interactions. In particular, they prove that a significant fraction of such a pre-accelerated particles is reflected upstream and further picked up by the Fermi I-order mechanism.

An interesting possibility arises in pulsar binary systems, where
the shocks may switch between different regimes -- when the
binary members are close, such that the distance between
the pulsar and the shock is smaller than the critical radius $r_{\rm shock}<r_{\rm crit}$ (\ref{propcond}),
the shock is in the MHD regime and the energy is dissipated via driven reconnection of stripes. 
When the separation between binary members becomes larger, such that the shock is located at a radius $r_{\rm shock}>r_{\rm crit}$, a shock can acquire an EM precursor. At this shock regime transition, a sudden appearance of a precursor ahead of the shock should be
accompanied by enhanced emission in the synchrotron and IC (on the photons from the stellar companion) processes, because the
precursor emission is the most efficient close to the cut-off \cite{2013ApJ...776...40M}.
This scenario has been invoked in the eccentric gamma-ray binary B1259-63 \cite{2013ApJ...776...40M}, where a mysterious flare, observed by the Fermi-LAT \cite{2011ApJ...736L..11A,2011ApJ...736L..10T,2015ApJ...798L..26T}, challenges the models \cite{2012ApJ...753..127K,2012ApJ...752L..17K,2013A&A...557A.127D}. 

\section{Summary}

Pulsar winds play an important role in the coupling between the central engine -- a rotating magnetized star -- and the nebula, downstream of the pulsar wind shock, where the electromagnetic energy carried by the wind is released into the plasma.  
The mechanism leading to the energy release is not clear, however. It seems plausible that it is related to the particular structure of the equatorial component of the wind, which has a form of stripes of opposite magnetic polarity, separated by the current sheet that performs a wavy pattern in space as the pulsar rotates. This structure is believed to be prone to the relativistic magnetic reconnection. 

Although it has been shown that the reconnection in the wind is not able to dissipate the electromagnetic energy before it reaches the termination shock, it provides a promising scenario for the production of the high-energy pulsed emission. The particles accelerated in the course of reconnection emit synchrotron and SSC photons, whose spectrum can reach TeV energies. Remarkably, recent observations of 
the Crab \cite{2016A&A...585A.133A}, as well as the millisecond pulsar J0614--3329 \cite{2016arXiv160408710X} confirm the detection of the very high energy spectral components, favouring the scenario, in which the gamma-rays originate from outside the light cylinder.

Most of the wind electromagnetic energy must be, however, released somewhere close to the termination shock of the wind. In this case, the stripes become compressed when the wind impacts on the shock, and the driven reconnection annihilates the alternating field component. The nonthermal spectrum of the particles accelerated in the course of this process can be observed, if the plasma carried by the wind is sufficiently dense. 
In the low plasma density, on the other hand, the electromagnetic fields of the wind behave more like
an electromagnetic wave instead of a familiar MHD wind. Recent studies show that the shocks in diluted plasmas exhibit dissipative, electromagnetic precursors, which mediate the Poynting flux dissipation and pre-accelerate particles, which are further a subject to the Fermi process. This scenario is expected to be realised in shocks of isolated pulsars, located at large stand-off distances, $r_{\rm shock}>\kappa\rl$.

\begin{acknowledgement}
I thank Tomasz Rembiasz and J\'er\^{o}me  P\'etri for helpful comments on the manuscript.
\end{acknowledgement} 

%
%
%
%



\begin{thebibliography}{100}

\bibitem{2010ApJ...708.1254A}
A.~A. {Abdo et al.}
\newblock {Fermi Large Area Telescope Observations of the Crab Pulsar And
  Nebula}.
\newblock {\em \apj}, 708:1254--1267, January 2010.

\bibitem{2011ApJ...736L..11A}
A.~A. {Abdo et al.}
\newblock {Discovery of High-energy Gamma-ray Emission from the Binary System
  PSR B1259-63/LS 2883 around Periastron with Fermi}.
\newblock {\em \apjl}, 736:L11, July 2011.

\bibitem{2014A&A...565L..12A}
J.~{Aleksi{\'c} et al.}
\newblock {Detection of bridge emission above 50 GeV from the Crab pulsar with
  the MAGIC telescopes}.
\newblock {\em \aap}, 565:L12, May 2014.

\bibitem{2013ApJ...770...18A}
T.~{Amano} and J.~G. {Kirk}.
\newblock {The Role of Superluminal Electromagnetic Waves in Pulsar Wind
  Termination Shocks}.
\newblock {\em \apj}, 770:18, June 2013.

\bibitem{2016A&A...585A.133A}
S.~{Ansoldi et al.}
\newblock {Teraelectronvolt pulsed emission from the Crab Pulsar detected by
  MAGIC}.
\newblock {\em \aap}, 585:A133, January 2016.

\bibitem{2012ApJ...745..108A}
I.~{Arka} and J.~G. {Kirk}.
\newblock {Superluminal Waves in Pulsar Winds}.
\newblock {\em \apj}, 745:108, February 2012.

\bibitem{1979ApJ...231..854A}
J.~{Arons} and E.~T. {Scharlemann}.
\newblock {Pair formation above pulsar polar caps - Structure of the low
  altitude acceleration zone}.
\newblock {\em \apj}, 231:854--879, August 1979.

\bibitem{1984A&A...139..417A}
E.~{Asseo}, R.~{Pellat}, and X.~{Llobet}.
\newblock {Spherical propagation of large amplitude pulsar waves}.
\newblock {\em \aap}, 139:417--425, October 1984.

\bibitem{2010ApJ...715.1270B}
X.-N. {Bai} and A.~{Spitkovsky}.
\newblock {Uncertainties of Modeling Gamma-ray Pulsar Light Curves Using Vacuum
  Dipole Magnetic Field}.
\newblock {\em \apj}, 715:1270--1281, June 2010.

\bibitem{1982ApJ...254..713B}
J.~J. {Barnard} and J.~{Arons}.
\newblock {Pair production and pulsar cutoff in magnetized neutron stars with
  nondipolar magnetic geometry}.
\newblock {\em \apj}, 254:713--734, March 1982.

\bibitem{2007PhPl...14e6503B}
N.~{Bessho} and A.~{Bhattacharjee}.
\newblock {Fast collisionless reconnection in electron-positron plasma}.
\newblock {\em Physics of Plasmas}, 14(5):056503, May 2007.

\bibitem{1999A&A...349.1017B}
S.~V. {Bogovalov}.
\newblock {On the physics of cold MHD winds from oblique rotators}.
\newblock {\em \aap}, 349:1017--1026, September 1999.

\bibitem{2011MNRAS.410..381B}
N.~{Bucciantini}, J.~{Arons}, and E.~{Amato}.
\newblock {Modelling spectral evolution of pulsar wind nebulae inside supernova
  remnants}.
\newblock {\em \mnras}, 410:381--398, January 2011.

\bibitem{1977MNRAS.180..125B}
R.~{Buckley}.
\newblock {Pulsar magnetospheres with arbitrary geometry in the force-free
  approximation}.
\newblock {\em \mnras}, 180:125--140, July 1977.

\bibitem{2015MNRAS.448..606C}
B.~{Cerutti}, A.~{Philippov}, K.~{Parfrey}, and A.~{Spitkovsky}.
\newblock {Particle acceleration in axisymmetric pulsar current sheets}.
\newblock {\em \mnras}, 448:606--619, March 2015.

\bibitem{2012ApJ...746..148C}
B.~{Cerutti}, D.~A. {Uzdensky}, and M.~C. {Begelman}.
\newblock {Extreme Particle Acceleration in Magnetic Reconnection Layers:
  Application to the Gamma-Ray Flares in the Crab Nebula}.
\newblock {\em \apj}, 746:148, February 2012.

\bibitem{2014ApJ...782..104C}
B.~{Cerutti}, G.~R. {Werner}, D.~A. {Uzdensky}, and M.~C. {Begelman}.
\newblock {Three-dimensional Relativistic Pair Plasma Reconnection with
  Radiative Feedback in the Crab Nebula}.
\newblock {\em \apj}, 782:104, February 2014.

\bibitem{1986ApJ...300..500C}
K.~S. {Cheng}, C.~{Ho}, and M.~{Ruderman}.
\newblock {Energetic radiation from rapidly spinning pulsars. I - Outer
  magnetosphere gaps. II - VELA and Crab}.
\newblock {\em \apj}, 300:500--539, January 1986.

\bibitem{1974JPlPh..12..297C}
P.~C. {Clemmow}.
\newblock {Nonlinear waves in a cold plasma by Lorentz transformation}.
\newblock {\em Journal of Plasma Physics}, 12:297--317, October 1974.

\bibitem{2002ApJ...566..336C}
I.~{Contopoulos} and D.~{Kazanas}.
\newblock {Toward Resolving the Crab {$\sigma$}-Problem: A Linear Accelerator?}
\newblock {\em \apj}, 566:336--342, February 2002.

\bibitem{1990ApJ...349..538C}
F.~V. {Coroniti}.
\newblock {Magnetically striped relativistic magnetohydrodynamic winds - The
  Crab Nebula revisited}.
\newblock {\em \apj}, 349:538--545, February 1990.

\bibitem{1982ApJ...252..337D}
J.~K. {Daugherty} and A.~K. {Harding}.
\newblock {Electromagnetic cascades in pulsars}.
\newblock {\em \apj}, 252:337--347, January 1982.

\bibitem{1996ApJ...457..253D}
O.~C. {de Jager}, A.~K. {Harding}, P.~F. {Michelson}, H.~I. {Nel}, P.~L.
  {Nolan}, P.~{Sreekumar}, and D.~J. {Thompson}.
\newblock {Gamma-Ray Observations of the Crab Nebula: A Study of the
  Synchro-Compton Spectrum}.
\newblock {\em \apj}, 457:253, January 1996.

\bibitem{2013APh....43..287D}
E.~{de O{\~n}a-Wilhelmi}, B.~{Rudak}, J.~A. {Barrio}, J.~L. {Contreras},
  Y.~{Gallant}, D.~{Hadasch}, T.~{Hassan}, M.~{Lopez}, D.~{Mazin},
  N.~{Mirabal}, G.~{Pedaletti}, M.~{Renaud}, R.~{de los Reyes}, D.~F. {Torres},
  and {CTA Consortium}.
\newblock {Prospects for observations of pulsars and pulsar wind nebulae with
  CTA}.
\newblock {\em Astroparticle Physics}, 43:287--300, March 2013.

\bibitem{2013A&A...557A.127D}
G.~{Dubus} and B.~{Cerutti}.
\newblock {What caused the GeV flare of PSR B1259-63?}
\newblock {\em \aap}, 557:A127, September 2013.

\bibitem{2003ApJ...598.1201D}
J.~{Dyks} and B.~{Rudak}.
\newblock {Two-Pole Caustic Model for High-Energy Light Curves of Pulsars}.
\newblock {\em \apj}, 598:1201--1206, December 2003.

\bibitem{2002ASPC..271...99G}
Y.~A. {Gallant}, E.~{van der Swaluw}, J.~G. {Kirk}, and A.~{Achterberg}.
\newblock {Modeling Plerion Spectra and their Evolution}.
\newblock In P.~O. {Slane} and B.~M. {Gaensler}, editors, {\em Neutron Stars in
  Supernova Remnants}, volume 271 of {\em Astronomical Society of the Pacific
  Conference Series}, page~99, 2002.

\bibitem{giacchekirk}
S.~{Giacche} and J.~G. Kirk.
\newblock {\em submitted}.

\bibitem{2001ApJ...550..383G}
J.~{Gil} and D.~{Mitra}.
\newblock {Vacuum Gaps in Pulsars and PSR J2144-3933}.
\newblock {\em \apj}, 550:383--391, March 2001.

\bibitem{2014PhRvL.113o5005G}
F.~{Guo}, H.~{Li}, W.~{Daughton}, and Y.-H. {Liu}.
\newblock {Formation of Hard Power Laws in the Energetic Particle Spectra
  Resulting from Relativistic Magnetic Reconnection}.
\newblock {\em Physical Review Letters}, 113(15):155005, October 2014.

\bibitem{1991Sci...251.1033H}
A.~K. {Harding}.
\newblock {Physics in strong magnetic fields near neutron stars}.
\newblock {\em Science}, 251:1033--1038, March 1991.

\bibitem{2015ApJ...811...63H}
A.~K. {Harding} and C.~{Kalapotharakos}.
\newblock {Synchrotron Self-Compton Emission from the Crab and Other Pulsars}.
\newblock {\em \apj}, 811:63, September 2015.

\bibitem{1978ApJ...225..226H}
A.~K. {Harding}, E.~{Tademaru}, and L.~W. {Esposito}.
\newblock {A curvature-radiation-pair-production model for gamma-ray pulsars}.
\newblock {\em \apj}, 225:226--236, October 1978.

\bibitem{2004PhPl...11.1151J}
C.~H. {Jaroschek}, R.~A. {Treumann}, H.~{Lesch}, and M.~{Scholer}.
\newblock {Fast reconnection in relativistic pair plasmas: Analysis of particle
  acceleration in self-consistent full particle simulations}.
\newblock {\em Physics of Plasmas}, 11:1151--1163, March 2004.

\bibitem{2015SSRv..191..545K}
D.~{Kagan}, L.~{Sironi}, B.~{Cerutti}, and D.~{Giannios}.
\newblock {Relativistic Magnetic Reconnection in Pair Plasmas and Its
  Astrophysical Applications}.
\newblock {\em \ssr}, 191:545--573, October 2015.

\bibitem{2012ApJ...749....2K}
C.~{Kalapotharakos}, D.~{Kazanas}, A.~{Harding}, and I.~{Contopoulos}.
\newblock {Toward a Realistic Pulsar Magnetosphere}.
\newblock {\em \apj}, 749:2, April 2012.

\bibitem{1984ApJ...283..694K}
C.~F. {Kennel} and F.~V. {Coroniti}.
\newblock {Confinement of the Crab pulsar's wind by its supernova remnant}.
\newblock {\em \apj}, 283:694--709, August 1984.

\bibitem{1984ApJ...283..710K}
C.~F. {Kennel} and F.~V. {Coroniti}.
\newblock {Magnetohydrodynamic model of Crab nebula radiation}.
\newblock {\em \apj}, 283:710--730, August 1984.

\bibitem{1976JPlPh..15..335K}
C.~F. {Kennel} and R.~{Pellat}.
\newblock {Relativistic nonlinear plasma waves in a magnetic field}.
\newblock {\em Journal of Plasma Physics}, 15:335--355, June 1976.

\bibitem{2012ApJ...752L..17K}
D.~{Khangulyan}, F.~A. {Aharonian}, S.~V. {Bogovalov}, and M.~{Rib{\'o}}.
\newblock {Post-periastron Gamma-Ray Flare from PSR B1259-63/LS 2883 as a
  Result of Comptonization of the Cold Pulsar Wind}.
\newblock {\em \apjl}, 752:L17, June 2012.

\bibitem{2004PhRvL..92r1101K}
J.~G. {Kirk}.
\newblock {Particle Acceleration in Relativistic Current Sheets}.
\newblock {\em Physical Review Letters}, 92(18):181101, May 2004.

\bibitem{2009ASSL..357..421K}
J.~G. {Kirk}, Y.~{Lyubarsky}, and J.~{Petri}.
\newblock {The Theory of Pulsar Winds and Nebulae}.
\newblock In W.~{Becker}, editor, {\em Astrophysics and Space Science Library},
  volume 357 of {\em Astrophysics and Space Science Library}, page 421, 2009.

\bibitem{2003ApJ...591..366K}
J.~G. {Kirk} and O.~{Skj{\ae}raasen}.
\newblock {Dissipation in Poynting-Flux-dominated Flows: The {$\sigma$}-Problem
  of the Crab Pulsar Wind}.
\newblock {\em \apj}, 591:366--379, July 2003.

\bibitem{2002A&A...388L..29K}
J.~G. {Kirk}, O.~{Skj{\ae}raasen}, and Y.~A. {Gallant}.
\newblock {Pulsed radiation from neutron star winds}.
\newblock {\em \aap}, 388:L29--L32, June 2002.

\bibitem{2006MNRAS.367...19K}
S.~S. {Komissarov}.
\newblock {Simulations of the axisymmetric magnetospheres of neutron stars}.
\newblock {\em \mnras}, 367:19--31, March 2006.

\bibitem{2012ApJ...753..127K}
S.~W. {Kong}, K.~S. {Cheng}, and Y.~F. {Huang}.
\newblock {Modeling the Multiwavelength Light Curves of PSR B1259-63/LS 2883.
  II. The Effects of Anisotropic Pulsar Wind and Doppler Boosting}.
\newblock {\em \apj}, 753:127, July 2012.

\bibitem{1985MNRAS.213P..43K}
J.~{Krause-Polstorff} and F.~C. {Michel}.
\newblock {Electrosphere of an aligned magnetized neutron star}.
\newblock {\em \mnras}, 213:43P--49P, March 1985.

\bibitem{1985A&A...144...72K}
J.~{Krause-Polstorff} and F.~C. {Michel}.
\newblock {Pulsar space charging}.
\newblock {\em \aap}, 144:72--80, March 1985.

\bibitem{2001A&A...378..918K}
L.~{Kuiper}, W.~{Hermsen}, G.~{Cusumano}, R.~{Diehl}, V.~{Sch{\"o}nfelder},
  A.~{Strong}, K.~{Bennett}, and M.~L. {McConnell}.
\newblock {The Crab pulsar in the 0.75-30 MeV range as seen by CGRO COMPTEL. A
  coherent high-energy picture from soft X-rays up to high-energy gamma-rays}.
\newblock {\em \aap}, 378:918--935, November 2001.

\bibitem{2014ApJ...797L..13L}
G.~C.~K. {Leung}, J.~{Takata}, C.~W. {Ng}, A.~K.~H. {Kong}, P.~H.~T. {Tam},
  C.~Y. {Hui}, and K.~S. {Cheng}.
\newblock {Fermi-LAT Detection of Pulsed Gamma-Rays above 50 GeV from the Vela
  Pulsar}.
\newblock {\em \apjl}, 797:L13, December 2014.

\bibitem{2012ApJ...746...60L}
J.~{Li}, A.~{Spitkovsky}, and A.~{Tchekhovskoy}.
\newblock {Resistive Solutions for Pulsar Magnetospheres}.
\newblock {\em \apj}, 746:60, February 2012.

\bibitem{1996A&A...311..172L}
Y.~E. {Lyubarskii}.
\newblock {A model for the energetic emission from pulsars.}
\newblock {\em \aap}, 311:172--178, July 1996.

\bibitem{2001ApJ...547..437L}
Y.~{Lyubarsky} and J.~G. {Kirk}.
\newblock {Reconnection in a Striped Pulsar Wind}.
\newblock {\em \apj}, 547:437--448, January 2001.

\bibitem{2003MNRAS.345..153L}
Y.~E. {Lyubarsky}.
\newblock {The termination shock in a striped pulsar wind}.
\newblock {\em \mnras}, 345:153--160, October 2003.

\bibitem{2005MNRAS.358..113L}
Y.~E. {Lyubarsky}.
\newblock {On the relativistic magnetic reconnection}.
\newblock {\em \mnras}, 358:113--119, March 2005.

\bibitem{2013MNRAS.431.2580L}
M.~{Lyutikov}.
\newblock {Inverse Compton model of pulsar high-energy emission}.
\newblock {\em \mnras}, 431:2580--2589, May 2013.

\bibitem{2003ApJ...589..893L}
M.~{Lyutikov} and D.~{Uzdensky}.
\newblock {Dynamics of Relativistic Reconnection}.
\newblock {\em \apj}, 589:893--901, June 2003.

\bibitem{1971PhRvL..27.1342M}
C.~{Max} and F.~{Perkins}.
\newblock {Strong Electromagnetic Waves in Overdense Plasmas}.
\newblock {\em Physical Review Letters}, 27:1342--1345, November 1971.

\bibitem{1996MNRAS.279.1168M}
A.~{Melatos} and D.~B. {Melrose}.
\newblock {Energy transport in a rotation-modulated pulsar wind.}
\newblock {\em \mnras}, 279:1168--1190, April 1996.

\bibitem{1973ApJ...180..207M}
F.~C. {Michel}.
\newblock {Rotating Magnetosphere: a Simple Relativistic Model}.
\newblock {\em \apj}, 180:207--226, February 1973.

\bibitem{1973ApJ...180L.133M}
F.~C. {Michel}.
\newblock {Rotating Magnetospheres: an Exact 3-D Solution}.
\newblock {\em \apjl}, 180:L133, March 1973.

\bibitem{1994ApJ...431..397M}
F.~C. {Michel}.
\newblock {Magnetic structure of pulsar winds}.
\newblock {\em \apj}, 431:397--401, August 1994.

\bibitem{2013ApJ...771...53M}
I.~{Mochol} and J.~G. {Kirk}.
\newblock {Propagation and Stability of Superluminal Waves in Pulsar Winds}.
\newblock {\em \apj}, 771:53, July 2013.

\bibitem{2013ApJ...776...40M}
I.~{Mochol} and J.~G. {Kirk}.
\newblock {Radiative Damping and Emission Signatures of Strong Superluminal
  Waves in Pulsar Winds}.
\newblock {\em \apj}, 776:40, October 2013.

\bibitem{2015MNRAS.449L..51M}
I.~{Mochol} and J.~{P{\'e}tri}.
\newblock {Very high energy emission as a probe of relativistic magnetic
  reconnection in pulsar winds}.
\newblock {\em \mnras}, 449:L51--L55, April 2015.

\bibitem{2003ApJ...588..430M}
A.~G. {Muslimov} and A.~K. {Harding}.
\newblock {Extended Acceleration in Slot Gaps and Pulsar High-Energy Emission}.
\newblock {\em \apj}, 588:430--440, May 2003.

\bibitem{2004ApJ...606.1143M}
A.~G. {Muslimov} and A.~K. {Harding}.
\newblock {High-Altitude Particle Acceleration and Radiation in Pulsar Slot
  Gaps}.
\newblock {\em \apj}, 606:1143--1153, May 2004.

\bibitem{2012MNRAS.424..605P}
J.~{P{\'e}tri}.
\newblock {The pulsar force-free magnetosphere linked to its striped wind:
  time-dependent pseudo-spectral simulations}.
\newblock {\em \mnras}, 424:605--619, July 2012.

\bibitem{2013MNRAS.434.2636P}
J.~{P{\'e}tri}.
\newblock {Phase-resolved polarization properties of the pulsar striped wind
  synchrotron emission}.
\newblock {\em \mnras}, 434:2636--2644, September 2013.

\bibitem{2015MNRAS.450..714P}
J.~{P{\'e}tri}.
\newblock {Multipolar electromagnetic fields around neutron stars: exact vacuum
  solutions and related properties}.
\newblock {\em \mnras}, 450:714--742, June 2015.

\bibitem{2016MNRAS.455.3779P}
J.~{P{\'e}tri}.
\newblock {General-relativistic force-free pulsar magnetospheres}.
\newblock {\em \mnras}, 455:3779--3805, February 2016.

\bibitem{2016arXiv160804895P}
J.~{P{\'e}tri}.
\newblock {Theory of pulsar magnetosphere and wind}.
\newblock {\em ArXiv e-prints}, August 2016.

\bibitem{2002A&A...384..414P}
J.~{P{\'e}tri}, J.~{Heyvaerts}, and S.~{Bonazzola}.
\newblock {Global static electrospheres of charged pulsars}.
\newblock {\em \aap}, 384:414--432, March 2002.

\bibitem{2005ApJ...627L..37P}
J.~{P{\'e}tri} and J.~G. {Kirk}.
\newblock {The Polarization of High-Energy Pulsar Radiation in the Striped Wind
  Model}.
\newblock {\em \apjl}, 627:L37--L40, July 2005.

\bibitem{2007A&A...473..683P}
J.~{P{\'e}tri} and Y.~{Lyubarsky}.
\newblock {Magnetic reconnection at the termination shock in a striped pulsar
  wind}.
\newblock {\em \aap}, 473:683--700, October 2007.

\bibitem{2015MNRAS.446.2243P}
S.~A. {Petrova}.
\newblock {Axisymmetric force-free magnetosphere of a pulsar - II. Transition
  from the self-consistent two-fluid model}.
\newblock {\em \mnras}, 446:2243--2250, January 2015.

\bibitem{2016arXiv160807998P}
S.~A. {Petrova}.
\newblock {A novel look at the pulsar force-free magnetosphere}.
\newblock {\em ArXiv e-prints}, August 2016.

\bibitem{2015ApJ...801L..19P}
A.~A. {Philippov}, A.~{Spitkovsky}, and B.~{Cerutti}.
\newblock {Ab Initio Pulsar Magnetosphere: Three-dimensional Particle-in-cell
  Simulations of Oblique Pulsars}.
\newblock {\em \apjl}, 801:L19, March 2015.

\bibitem{2013MNRAS.431L..48P}
O.~{Porth}, S.~S. {Komissarov}, and R.~{Keppens}.
\newblock {Solution to the sigma problem of pulsar wind nebulae}.
\newblock {\em \mnras}, 431:L48--L52, April 2013.

\bibitem{1974MNRAS.167....1R}
M.~J. {Rees} and J.~E. {Gunn}.
\newblock {The origin of the magnetic field and relativistic particles in the
  Crab Nebula}.
\newblock {\em \mnras}, 167:1--12, April 1974.

\bibitem{1995ApJ...438..314R}
R.~W. {Romani} and I.-A. {Yadigaroglu}.
\newblock {Gamma-ray pulsars: Emission zones and viewing geometries}.
\newblock {\em \apj}, 438:314--321, January 1995.

\bibitem{velahess2}
{Rudak et al. for the HESS II Collaboration, B.}
\newblock {Pulsations from the Vela Pulsar Down to 20 GeV with H.E.S.S. II}.
\newblock 2015.

\bibitem{1975ApJ...196...51R}
M.~A. {Ruderman} and P.~G. {Sutherland}.
\newblock {Theory of pulsars - Polar caps, sparks, and coherent microwave
  radiation}.
\newblock {\em \apj}, 196:51--72, February 1975.

\bibitem{1973ApJ...182..951S}
E.~T. {Scharlemann} and R.~V. {Wagoner}.
\newblock {Aligned Rotating Magnetospheres. General Analysis}.
\newblock {\em \apj}, 182:951--960, June 1973.

\bibitem{2016MNRAS.462...48S}
L.~{Sironi}, D.~{Giannios}, and M.~{Petropoulou}.
\newblock {Plasmoids in relativistic reconnection, from birth to adulthood:
  first they grow, then they go}.
\newblock {\em \mnras}, 462:48--74, October 2016.

\bibitem{2011ApJ...741...39S}
L.~{Sironi} and A.~{Spitkovsky}.
\newblock {Acceleration of Particles at the Termination Shock of a Relativistic
  Striped Wind}.
\newblock {\em \apj}, 741:39, November 2011.

\bibitem{2014ApJ...783L..21S}
L.~{Sironi} and A.~{Spitkovsky}.
\newblock {Relativistic Reconnection: An Efficient Source of Non-thermal
  Particles}.
\newblock {\em \apjl}, 783:L21, March 2014.

\bibitem{2004IAUS..218..357S}
A.~{Spitkovsky}.
\newblock {Electrodynamics of Pulsar Magnetospheres}.
\newblock In F.~{Camilo} and B.~M. {Gaensler}, editors, {\em Young Neutron
  Stars and Their Environments}, volume 218 of {\em IAU Symposium}, page 357,
  2004.

\bibitem{2006ApJ...648L..51S}
A.~{Spitkovsky}.
\newblock {Time-dependent Force-free Pulsar Magnetospheres: Axisymmetric and
  Oblique Rotators}.
\newblock {\em \apjl}, 648:L51--L54, September 2006.

\bibitem{1971ApJ...164..529S}
P.~A. {Sturrock}.
\newblock {A Model of Pulsars}.
\newblock {\em \apj}, 164:529, March 1971.

\bibitem{2011ApJ...736L..10T}
P.~H.~T. {Tam}, R.~H.~H. {Huang}, J.~{Takata}, C.~Y. {Hui}, A.~K.~H. {Kong},
  and K.~S. {Cheng}.
\newblock {Discovery of GeV {$\gamma$}-ray Emission from PSR B1259-63/LS 2883}.
\newblock {\em \apjl}, 736:L10, July 2011.

\bibitem{2015ApJ...798L..26T}
P.~H.~T. {Tam}, K.~L. {Li}, J.~{Takata}, A.~T. {Okazaki}, C.~Y. {Hui}, and
  A.~K.~H. {Kong}.
\newblock {High-energy Observations of PSR B1259-63/LS 2883 through the 2014
  Periastron Passage: Connecting X-Rays to the GeV Flare}.
\newblock {\em \apjl}, 798:L26, January 2015.

\bibitem{2013MNRAS.435L...1T}
A.~{Tchekhovskoy}, A.~{Spitkovsky}, and J.~G. {Li}.
\newblock {Time-dependent 3D magnetohydrodynamic pulsar magnetospheres: oblique
  rotators}.
\newblock {\em \mnras}, 435:L1--L5, August 2013.

\bibitem{2015ApJ...810..144T}
A.~N. {Timokhin} and A.~K. {Harding}.
\newblock {On the Polar Cap Cascade Pair Multiplicity of Young Pulsars}.
\newblock {\em \apj}, 810:144, September 2015.

\bibitem{1975Ap&SS..32..375U}
V.~V. {Usov}.
\newblock {Wave zone structure of NP 0532 and infrared radiation excess of Crab
  Nebula}.
\newblock {\em \apss}, 32:375--377, February 1975.

\bibitem{2015MNRAS.449.3755V}
D.~{Vigan{\`o}} and D.~F. {Torres}.
\newblock {Modelling of the {$\gamma$}-ray pulsed spectra of Geminga, Crab, and
  Vela with synchro-curvature radiation}.
\newblock {\em \mnras}, 449:3755--3765, June 2015.

\bibitem{2000ApJ...536L..81W}
M.~C. {Weisskopf}, J.~J. {Hester}, A.~F. {Tennant}, R.~F. {Elsner}, N.~S.
  {Schulz}, H.~L. {Marshall}, M.~{Karovska}, J.~S. {Nichols}, D.~A. {Swartz},
  J.~J. {Kolodziejczak}, and S.~L. {O'Dell}.
\newblock {Discovery of Spatial and Spectral Structure in the X-Ray Emission
  from the Crab Nebula}.
\newblock {\em \apjl}, 536:L81--L84, June 2000.

\bibitem{2016ApJ...816L...8W}
G.~R. {Werner}, D.~A. {Uzdensky}, B.~{Cerutti}, K.~{Nalewajko}, and M.~C.
  {Begelman}.
\newblock {The Extent of Power-law Energy Spectra in Collisionless Relativistic
  Magnetic Reconnection in Pair Plasmas}.
\newblock {\em \apjl}, 816:L8, January 2016.

\bibitem{2016arXiv160408710X}
Y.~{Xing} and Z.~{Wang}.
\newblock {Fermi Study of gamma-ray Millisecond Pulsars: the Spectral Shape and
  Pulsed 25--200 GeV Emission from J0614-3329}.
\newblock {\em ArXiv e-prints}, April 2016.

\bibitem{2001ApJ...562L..63Z}
S.~{Zenitani} and M.~{Hoshino}.
\newblock {The Generation of Nonthermal Particles in the Relativistic Magnetic
  Reconnection of Pair Plasmas}.
\newblock {\em \apjl}, 562:L63--L66, November 2001.

\end{thebibliography}

\backmatter


\end{document}